\documentclass[twocolumn]{aastex631}

\usepackage{color, soul}
\usepackage{etoolbox}
\makeatletter % Workaround to only color the year for cite commands
 \patchcmd{\NAT@citex}
   {\@citea\NAT@hyper@{%
     \NAT@nmfmt{\NAT@nm}%
     \hyper@natlinkbreak{\NAT@aysep\NAT@spacechar}{\@citeb\@extra@b@citeb}%
     \NAT@date}}
   {\@citea\NAT@nmfmt{\NAT@nm}%
   \NAT@aysep\NAT@spacechar\NAT@hyper@{\NAT@date}}{}{}

 \patchcmd{\NAT@citex}
   {\@citea\NAT@hyper@{%
     \NAT@nmfmt{\NAT@nm}%
     \hyper@natlinkbreak{\NAT@spacechar\NAT@@open\if*#1*\else#1\NAT@spacechar\fi}%
       {\@citeb\@extra@b@citeb}%
     \NAT@date}}
   {\@citea\NAT@nmfmt{\NAT@nm}%
   \NAT@spacechar\NAT@@open\if*#1*\else#1\NAT@spacechar\fi\NAT@hyper@{\NAT@date}}
   {}{}
\makeatother

\newcommand{\Sedona}{\textsc{Sedona}}
\newcommand{\Athena}{\textsc{Athena++}}

\newtoggle{referee}            %
\togglefalse{referee}          %

\iftoggle{referee}{%           %
  \usepackage{color,soul}    %
  \soulregister\citet7
  \soulregister\citep7
  \soulregister\ref7
}{%                            %
}                              %
\usepackage{xcolor}
\usepackage{amsmath}
\usepackage{CJK}
\usepackage{setspace}
\usepackage{cprotect}

\received{September 27, 2022}
\revised{ ??? }
\accepted{ ??? }
\submitjournal{ApJS}

\shorttitle{Synthesizing Spectra from 3D RHD Models}
\shortauthors{Schultz et al.}

\graphicspath{{./}{figures/}}
\begin{document}
\begin{CJK*}{UTF8}{gbsn}

\title{Synthesizing Spectra from 3D Radiation Hydrodynamic Models of Massive Stars Using Monte Carlo Radiation Transport \footnote{Released on ??, ??, 2022}}
%%%%%%%%%%%%%%%%%%%%%%%%%%%%%%%%%%%%%%%%%%\correspondingauthor{W. C. Schultz}
\email{wcschultz@physics.ucsb.edu}

\author[0000-0003-1796-9849]{William C. Schultz}
\affiliation{Department of Physics, University of California, Santa Barbara, CA 93106, USA}

\author[0000-0002-6543-2993]{Benny T.-H. Tsang}
\affiliation{Department of Astronomy and Theoretical Astrophysics Center, University of California, Berkeley, CA 94720, USA}

\author[0000-0001-8038-6836]{Lars Bildsten}
\affiliation{Department of Physics, University of California, Santa Barbara, CA 93106, USA}
\affiliation{Kavli Institute for Theoretical Physics, University of California, Santa Barbara, CA 93106, USA}

\author[0000-0002-2624-3399]{Yan-Fei Jiang (姜燕飞)}
\affiliation{Center for Computational Astrophysics, Flatiron Institute, New York, NY 10010, USA}

\begin{abstract}
Observations indicate that turbulent motions are present on most massive star surfaces. 
Starting from the observed phenomena of spectral lines with widths much larger than thermal broadening (e.g. micro- and macroturbulence) to the detection of stochastic low-frequency variability (SLFV) in the Transiting Exoplanet Survey Satellite photometry, these stars clearly have large scale turbulent motions on their surfaces. 
The cause of this turbulence is debated, with near-surface convection zones, core internal gravity waves, and wind variability being proposed. 
Our 3D grey radiation hydrodynamic (RHD) models characterized the surfaces’ convective dynamics driven by near-surface convection zones and provided a reasonable match to the observed SLFV in the most luminous massive stars. 
We now explore the complex emitting surfaces of these 3D RHD models, which strongly violate the 1D assumption of a plane parallel atmosphere.
By post-processing the grey RHD models with the Monte Carlo radiation transport code \Sedona, we synthesize stellar spectra and extract information from the broadening of individual photospheric lines. 
The use of \Sedona\ enables the calculation of the viewing angle and temporal dependence of spectral absorption line profiles. 
Combining uncorrelated temporal snapshots together, we compare the broadening from the 3D RHD models' velocity fields to the thermal broadening of the extended emitting region, showing that our synthesized spectral lines closely resemble the observed macroturbulent broadening from similarly luminous stars. 
More generally, the new techniques we have developed will allow for systematic studies of the origin of turbulent velocity broadening from any future 3D simulations.  
\end{abstract}



\section{Introduction} \label{sec:intro}
Observations of massive stars exhibit clear signs of highly turbulent surfaces.
The Transiting Exoplanet Survey Satellite \citep[TESS,][]{Ricker2015} has observed many O- and B- stars and found that nearly all exhibit brightening variations at low frequencies, $\nu \lesssim 10^{-4}\,$Hz, or 10 cycles per day \citep{Bowman2019b, Burssens2020}.
Termed stochastic low-frequency variability (SLFV), it is thought to be produces by large scale turbulent motions.
The physical mechanism that generates this SLFV could have several origins: turbulent plumes from near-surface convection zones \citep{Cantiello2021, Schultz2022}, internal gravity waves (IGWs) launched from the convective core \citep{Aerts2015, Edelmann2019}, and wind variability and dynamics \citep{Krticka2021}.
Though near-surface convection zones appear to generate the same trend as the observed SLFV strength and characteristic frequency \citep{Cantiello2021, Schultz2022}, the question still remains worthy of exploration across the Hertzsprung-Russell (HR) diagram.
Additionally, \cite{Schultz2022} found the photospheric velocities seen in 3D models are comparable to the observed spectroscopic broadening of similar stars.

Analyzing spectral lines from these massive stars is a distinct additional probe, as photospheric spectral lines are broadened by the velocity fields in the emitting regions.
Analysis of 2D models by\cite{Rogers2013} suggested that IGWs could generate photospheric spectral line variability that could be observable by future surveys \citep{Aerts2015}.
However, full 3D frequency dependent radiation hydrodynamics (RHD) models have not been computed, and are a challenge with current computing power.

To understand how spectral line broadening provides local velocity information, we look to previous methods of modeling lower mass stellar spectra \citep{Gray2005}.
Typically there are four main velocities used to fit photospheric spectral lines.
Thermal broadening comes from the intrinsic Maxwell-Boltzmann velocity distribution of the ions, $v_{\rm therm}$, and is Gaussian in profile.
This is often combined with the intrinsic broadening, arising from the atomic physics governing the line level transitions, to generate a Voigt profile for the spectral line. 
Projected rotational broadening, $v\sin i$, imparts a deep, steep walled trench shape on stellar spectral lines as half the star is red-shifted while the other half is blue-shifted.

The other two velocities arise from the impact of turbulent motions in the line forming regions.
The microturbulent velocity, denoted as $\xi$, is defined as an additional velocity impacting scales smaller than the emitting region and is added in quadrature to $v_{\rm therm}$ in the Gaussian broadening of spectral lines.
As $\xi$ affects the equivalent width, it is typically quantified using Curve of Growth analyses of heavy elements for which $\xi \gg v_{\rm therm}$.
The other turbulent broadening arises from macroturbulence, $v_{\rm macro}$, which corresponds to dynamics on scales larger than the emitting region.
This broadens the wings of the spectral lines and is typically used as a fitting parameter as discussed in \cite{Gray2005}.

Though these velocity choices were inherited from investigations of lower mass stars, they are very effective in fitting hot, massive star spectral lines \citep{Simon-Diaz2010, SimonDiaz2014, Simon-Diaz2017, SimonDiaz2018, Holgado2022}.
However because the turbulent surfaces of massive stars are significantly different than their low mass counterparts \citep{Jiang2018, Schultz2020, Schultz2022}, it is uncertain just how accurate the inferred $\xi$ and $v_{\rm macro}$ will be based on the existing line fitting approach.
One example of an unexplained discrepancy is a strong positive correlation between $v\sin i$ and $v_{\rm macro}$ for massive stars ($M>20\,M_{\odot}$) across the main sequence \citep{Simon-Diaz2017}.
Additionally when using the current fitting method, larger $\xi$ appears to limit the range of recoverable $v\sin i$ and $v_{\rm macro}$ \citep{SimonDiaz2014} implying that a quantification of these broadening velocities as well as a better theoretical understanding of their meaning in turbulent massive star surfaces are needed. 

Additionally, the violent, turbulent plume structures on the surfaces of these massive stars suggest that 1D spectral synthesis models \citep[e.g. \textsc{FASTWIND},][]{Santolaya-Rey1997} may be insufficient.
A more realistic method to probe their chaotic surface dynamics are 3D RHD models where the interplay between radiation and matter is taken into account in the excitation of the turbulence.
Similar investigations \citep[see,][and references there-in]{Dravins2021} have been performed for lower mass (F,G,K) stars where gas pressure dominates radiation and the velocities are subsonic, allowing 3D hydrodynamic models to be used without the need for radiation.
These hydro models were then post-processed with radiation transport to synthesize stellar spectra with good agreement.
Similar work with RHD models has been done for red supergiants (RSG) where radiation becomes an important source of pressure support near the surface \citep{Chiavassa2011a}.
Using the 3D radiation transport code, OPTIM3D, \cite{Chiavassa2009} post-processed their CO$^5$BOLD \citep{Freytag2002, Freytag2008} RHD models, using tabulated extinction coefficients specific to $T$ and $\rho$ of RSG surfaces to generate wide frequency range spectra.

For hotter massive star envelopes, the problem requires more computational resources as the entire envelopes are supported by the interplay between plasma, radiation, and turbulence.
Near-surface convection zones excited by near-Eddington luminosities traveling through opacity peaks lead to turbulent dynamics and inefficient convection as both the motions become trans-sonic and the optical depths are low enough that rising plumes can lose heat from radiative cooling \citep{Goldberg2022, Schultz2022}.
Fortunately, this physics is within the reach of our 3D grey RHD simulations, which have been run using \Athena\ \citep{Stone2020}, and we can now post-process them for the desired spectroscopic analysis.

We chose to post-process our \Athena\ models using \Sedona, a Monte Carlo (MC) radiation transport framework originally developed to model supernova light curves, spectra, and polarization \citep{Kasen2006}.
\Sedona\ calculates frequency-dependent opacities directly from atomic data in the co-moving frame of each cell without the need for the Sobolev or line expansion approximation, allowing for generic applications to a wide range of stellar envelopes with adjustable frequency resolution.
Additionally, the MC transport is beneficial as it directly tracks the transport paths of photons through the clumpy and dynamic plume structures and can be used to estimate viewing angle dependencies from 3D simulations.
Though not used in this analysis, \Sedona\ allows for future expansions of this work: exploring polarization of the emitted flux, comparing grey and non-grey radiation transport, and comparing LTE and non-LTE calculations in investigations of wind lines.
Two major modifications to \Sedona\ were needed to adapt it to massive star surface.
A 3D spherical wedge geometry, matching that of our \Athena\ simulations, was added to be able to initialize our models in the \Sedona\ framework.
Additionally, a novel emission method was developed to avoid transporting MC particles through excessively optically thick zones.
Detailed discussion of these modifications are presented in \S~\ref{sec:rad_trans}.

Generating synthetic spectral lines from a turbulent stellar surface is nontrivial.
The turbulence is excited by forests of lines and ionization changes that make up the Fe and He opacity peaks increasing the required frequency resolution and greatly increasing the effective optical depths compared to the grey case.
The requisite large frequency range as well as the high resolution of the simulation make these runs computationally expensive.
Additionally the near Eddington limited radiation field and $\rho$ fluctuations from the trans-sonic turbulence combine to launch plumes from the stellar surface creating a dynamic emitting region for the spectral lines that are starkly different from the 1D atmospheric models.
This strong turbulence also generates large velocity contrasts, broadening the lines ($\Delta \lambda \sim 5\,$\AA) to create line blending not expected in static models.

In this work we present our method for post-processing 3D models to synthesize stellar spectra.
The 3D grey RHD simulations used for post-processing are described in \S~\ref{sec:grey_models}, specifically presenting model parameters for the three models investigated and a comparison of the photosphere versus $\tau=1$ surfaces in the models.
In \S~\ref{sec:rad_trans} we detail the Monte Carlo radiation transport methods and compare grey transport between the simulations.
We describe the process of synthesizing spectral lines in \S~\ref{sec:post-process} as well as depicting the variety in emitting regions for individual lines in the models.
In \S~\ref{sec:model_spec} we highlight the results from one spectral line from each model and compare the impacts of $v_{\rm therm}$ and the models' velocity fields, predicting implications for $\xi$ and $v_{\rm macro}$.
We conclude with a brief summary of future work these new methods will allow in \S~\ref{sec:conclusion}.

\section{3D Grey Radiation Hydrodynamic Modeling } \label{sec:grey_models}

A number of 3D grey radiation hydrodynamic (RHD) models have simulated near-surface convection zones in massive star envelopes \citep{Jiang2015, Jiang2018, Schultz2022}.
These models were performed using \Athena\ \citep{Stone2020} and with additional radiation transport \citep{Jiang2012, Jiang2014, Jiang2021}.
In this work, we showcase our novel spectral synthesis software pipeline by post-processing three specific \Athena\ models from the simulation suite.
In Section \ref{subsec:Athena_models}, we briefly summarize the numerical treatments and the stellar properties of the envelope models.
We then explain how the common interpretation of having the photosphere at a constant radius is incomplete (Section \ref{subsec:photosphere}), and introduce the thermalization optical depth as a reliable basis for estimating where spectral features are formed and explore what these surfaces look like in our 3D models (Section \ref{sec:tau_th}).

\subsection{\Athena\ Model Description} \label{subsec:Athena_models}
The \Athena\ stellar envelope models were run on spherical-polar grids and used an HLLC Riemann solver for the hydrodynamics.
The radiation transport equation is solved implicitly using discrete ordinates and assuming local thermal equilibrium (LTE), following \cite{Jiang2021}.
In our model suite, the finite volume approach of radiation transport discretizes the radiation field of each cell into 120 angular directions over the full 4$\pi$ solid angle.
Gravitational acceleration follows $g \propto M_{\rm core}/r^{2}$, not including self-gravity of the envelope, as the envelope masses within the models, $M_{\rm env}$, are $<0.1\%$ of the core mass, $M_{\rm core}$.
Rotation and magnetic fields are omitted from these models as they are not expected to significantly affect the near-surface opacity peaks or the turbulence they excite \citep{Jiang2017, Cantiello2021, Schultz2022}.
As the ionization states of the plasma do not drastically affect the equation of state in stellar atmospheres with $T_{\rm eff} \gtrsim 10,000\,$K, constant solar mean molecular weights are used in these models.
These models utilize grey OPAL Rosseland mean opacities \citep{Iglesias1996} at solar composition in a fully ionized plasma.
The grey opacity approximation and LTE assumption likely break down outside the stellar surface where the continuum opacity falls below atomic line opacity.
Thus our analysis is limited to photospheric spectral lines.

\begin{splitdeluxetable*}{lcccccccccBlccccccccccc}
\tabletypesize{\footnotesize}
\tablenum{1}
\tablecaption{Properties of the 3D Stellar Models \label{tab:3D_models}}
\tablewidth{0pt}
\tablehead{
\colhead{Name} & \multicolumn{2}{c}{Masses} & \multicolumn{2}{c}{Temperature} & \multicolumn{2}{c}{Luminosities} & \colhead{Resolution} & \multicolumn{2}{c}{Angular Size} & \colhead{Name} & \multicolumn{5}{c}{Radii} & \multicolumn{2}{c}{Optical Depth} & \multicolumn{2}{c}{RMS Velocities} & \multicolumn{2}{c}{Scale Heights} \\
\nocolhead{Name} & \colhead{$M_{\rm core}$} & \colhead{$M_{\rm env}$} & \colhead{$T_{\mathrm{eff},F_{\rm r}}$ $^{\rm a}$} & \colhead{$T_{\rm eff}$ $^{\rm b}$} & \colhead{$L$} & \colhead{$L_{\rm Edd}$ $^{\rm c}$} & \colhead{$n_{\rm r}$, $n_{\theta}$, $n_{\phi}$} & \colhead{$\theta_{\rm min}, \theta_{\rm max}$ $|$ $\phi_{\rm min}, \phi_{\rm max}$} & \colhead{$\Omega_{\rm sim}$} & \nocolhead{Name} & \colhead{$r_{\rm base}$} & \colhead{$r_{\rm Fe }$} & \colhead{$r_{F_{\rm r} = \sigma T^4}$ $^{\rm d}$} & \colhead{$r_{\tau = 1}$ $^{\rm e}$} & \colhead{$r_{\rm max}$} & \colhead{$\tau_{\rm Fe}$} & \colhead{$\frac{\tau_{\rm Fe}}{\tau_{\rm crit}}$} & \colhead{$v_{\mathrm{r},\, \tau=1}$} & \colhead{$v_{\bot,\, \tau=1}$} & \colhead{$H_{\rm Fe}$} & \colhead{$H_{\tau=1}$} \\
\nocolhead{Name} & \colhead{(M$_{\odot}$)} & \colhead{(M$_{\odot}$)} & \colhead{($10^3\,$K)} & \colhead{($10^3\,$K)} & \colhead{(log($L$/L$_{\odot}$))} & \colhead{(log($L$/L$_{\odot}$))} & \nocolhead{res} & \colhead{($\pi,\pi$ $|$ $\pi,\pi$)} & \colhead{($\mathrm{sr}/\pi$)} & \nocolhead{Name} & \colhead{(R$_{\odot}$)} & \colhead{(R$_{\odot}$)} & \colhead{(R$_{\odot}$)} & \colhead{(R$_{\odot}$)} & \colhead{(R$_{\odot}$)} & \nocolhead{tau} & \nocolhead{tau} & \colhead{(km$\,$s$^{-1}$)} & \colhead{(km$\,$s$^{-1}$)} & \colhead{(R$_{\odot}$)}  & \colhead{(R$_{\odot}$)}
}
%\decimalcolnumbers
\startdata
T19L6.4 & 80 & 0.011 & 25 & 19 & 6.4 & 6.42 & 512, 256, 512 & $0.25,0.75$ $|$ $0,1$ & 1.41 & T19L6.4 & 16.3 & 44.0 & 78.6 & 99.1 & 335.5 & 5,200 & 2.5 & 120 & 95 & 4.36 & 13.2 \\
T32L5.2 & 35 & $10^{-4}$ & 35 & 32 & 5.2 & 6.07 & 336, 256, 256 & $0.44,0.56$ $|$ $0,0.11$ & 0.04 & T32L5.2 & 9.7 & 12.5 & 13.07 & 13.15 & 15.3 & 1,182 & 0.28 & 68 & 78 & 0.24 & 0.09 \\
T42L5.0 & 35 & $10^{-6}$ & 42 & 42 & 5.0 & 6.07 & 384, 128, 128 & $0.49,0.51$ $|$ $0,0.02$ & 0.0012 & T42L5.0 & 6.8 & 7.7 & 7.86 & 7.86 & 8.2 & 490 & 0.02 & 2.5 & 8.6 & 0.07 & 0.01\\
\enddata

\tablenotetext{\rm a}{ Defined as the angle-averaged temperature at $r_{F_{\rm r} = \sigma T^4}$ ($\langle T (r_{F_{\rm r} = \sigma T^4}) \rangle$).}
\tablenotetext{\rm b}{ Defined as the angle-averaged temperature at $r_{\tau = 1}$.}
\tablenotetext{\rm c}{ For an assumed electron scattering opacity.}
\tablenotetext{\rm d}{ Defined as the radius where the angle-averaged radiative flux is follows the Stefan-Boltzmann law \\ ($\langle F_{\rm r}(r_{F_{\rm r} = \sigma T^4})\rangle = \sigma \langle T(r_{F_{\rm r} = \sigma T^4}) \rangle^4$).}
\tablenotetext{\rm e}{ Defined as the radius where the angle-averaged optical depth is unity ($\langle \tau (r_{\tau =1}) \rangle = 1$).}
\tablenotetext{*}{ Note that T32L5.2 is the same model as T35L5.2 in \cite{Schultz2022} as a different $T_{\rm eff}$ definition was used.}
\end{splitdeluxetable*}

The \Athena\ stellar envelope model parameters for the simulations used here are summarized in Table \ref{tab:3D_models}.
The models are named by their $T_{\rm eff}/10^3\,$K and $\log(L/L_{\odot})$, and simulate the outer convective regions between boundary radii of $r_{\rm base}$ to $r_{\rm max}$.
Because the pressure scale height ($H \equiv P / \rho g$) is much smaller than the radial extent in these stars ($H \ll R$ in Table~\ref{tab:3D_models}), only a small spherical wedge is modeled over a limited solid angle ($\Omega_{\rm sim} < 4\pi$) to limit the computational costs.
The size of each wedge was chosen such that it would include at least $10$ scale heights in the angular dimensions at the location of the Fe opacity peak, i.e., $(\phi_{\rm max}-\phi_{\rm min})\times r_{\rm Fe} \geq 10\,H_{\rm Fe}$.
The radial refinement is set to $\delta r / r \lesssim 0.1\%$ and the angular resolution ensures $r\delta \theta \approx r\delta\phi \approx \delta r$.
This allows each model to capture $\approx 5$ convective plumes, each more than 30 model cells across.
The $\theta$ extent of each simulation domain is centered on $\pi$ to alleviate numerical problems at the poles.
The inner boundary condition fixes $T, \rho, F_{\rm r}$, and $v_{\rm r}=v_{\bot}=0$ in the radiative region between the core and modeled regions while the outer boundary is open to allow radiation to escape and possible outflows.
The angular boundaries along the $\theta$ and $\phi$ directions are periodic.

The \Athena\ models are initialized using 1D models from the Modules for Experiments in Stellar Astrophysics (MESA) \citep{Paxton2011,Paxton2013,Paxton2015,Paxton2018,Paxton2019,Jermyn2022} using the default inlists.
The $T$, $\rho$, and $r$ at the Fe opacity peak in the MESA models are used with the $M$ and $L$ to create a hydrostatic model that is in thermal equilibrium without any convection, which initializes the 3D simulations.
The outer regions of the simulation, outside of the stellar profile, are initialized with a low density floor ($\rho = 10^{-17}$g$\,$cm$^{-3}$) to allow the models to find equilibrium without affecting the envelope mass and results in an integrated optical depth, $\tau_{\rm floor} \approx 10^{-6}$.
When the \Athena\ run begins, the initial solution is unstable to convective motion and the near-surface convection zones develop.
The models are run until the time-averaged radial energy flux has converged and the Fe opacity peak has experienced at least three thermal times, $t_{\rm therm} = \int_{r_{\rm Fe}}^{r_{\rm max}} C_{\rm P} T / L dM$, where $C_{\rm P}$ is the heat capacity at constant pressure. 
Once these conditions have been met, typically after r $\gtrsim 10^7$ core hours, we are satisfied that steady-state equilibrium has been reached.

\begin{figure*}
%\vspace*{-2.0 cm}
\begin{center}
 \includegraphics[width=0.85\textwidth]{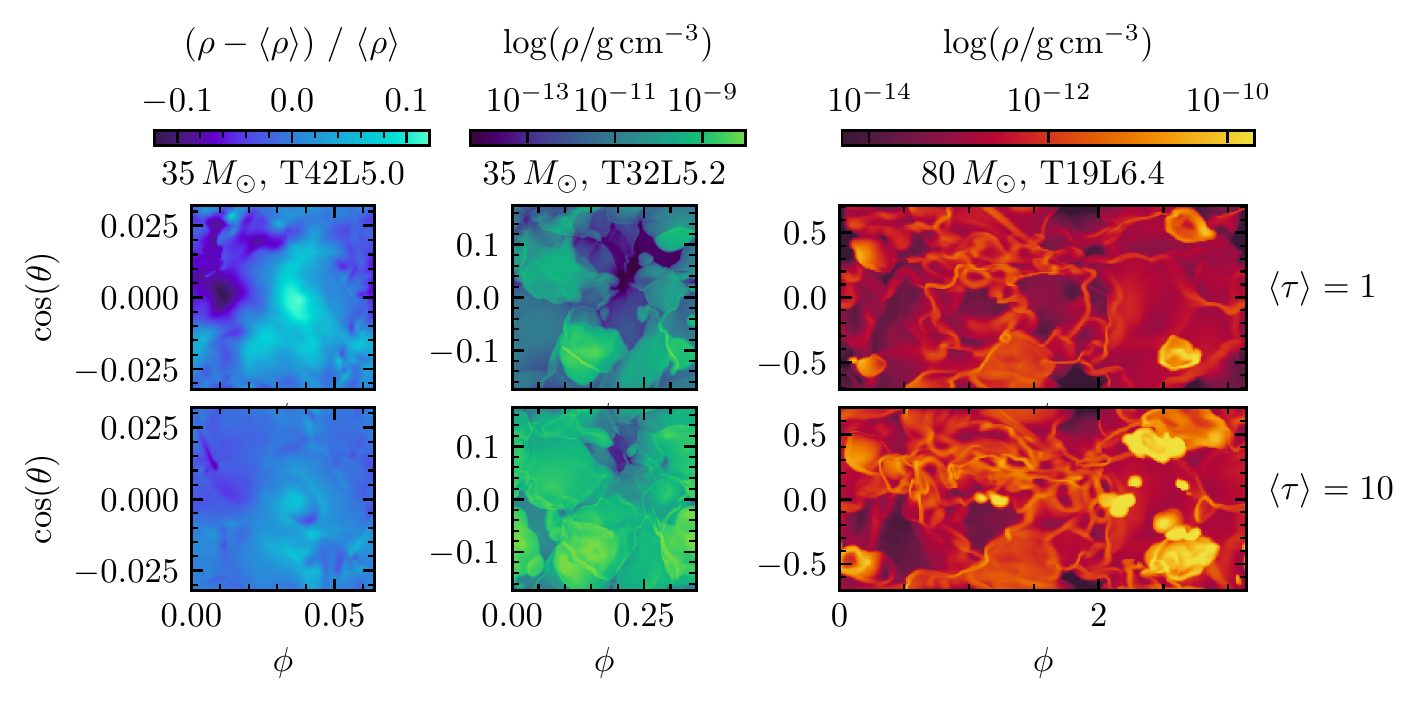} 
%\vspace*{-1.0 cm}
\caption{Density fluctuations through radial slices where the angle-averaged optical depth, $\langle \tau \rangle$, is 1 (top) and 10 (bottom) for T42L5.0, T32L5.2, and T19L6.4 (left to right).
Colors show over- and under-densities in the left column and $\log\rho$ for the middle and right.}
\label{fig:rslice}
\end{center}
\end{figure*}

All these models are hot, bright stars with T19L6.4 in the Hertzsprung Gap, T32L5.2 in the middle of Main Sequence evolution, and T42L5.0 near the Zero-Age Main Sequence.
Their luminosities are nearly Eddington limited for an electron scattering opacity, $L\sim L_{\rm Edd}\equiv 4\pi G M c / \kappa_{\rm es}$, and thus any increase in opacity from prevailing Fe and He opacity peaks will cause vigorous convective dynamics.
These convective zones are best understood by quantifying their convective efficiency using an optical depth diagnostic.
In optically thick convection zones, the convective efficiency can be written as the advective flux over the diffusive radiative flux, or approximately $\gamma = (P_{\rm r} + P_{\rm g}) v_{\rm conv} / P_{\rm r} (c/\tau)$ where $v_{\rm conv}$ is the convective velocity and $P_{\rm r}$ and $P_{\rm g}$ are the radiation and gas pressures respectively.
We solve for an optical depth at which $\gamma = 1$, or the point at which both transport methods are equal, to define $\tau_{\rm crit}$,
\begin{equation}
    \tau_{\rm crit} \equiv \frac{c P_{\rm r}}{v_{\rm RMS}(P_{\rm r} + P_{\rm g})},
\end{equation}
where we have assumed the convective velocity is similar to the RMS velocity, $v_{\rm conv} \approx v_{\rm RMS}$ \citep{Jiang2015,Schultz2020,Goldberg2022}.
Thus if a convective zone lies at $\tau > \tau_{\rm crit}$, we expect convective energy transport to dominate there.
Otherwise, though still convective, the majority of the luminosity is carried via radiative diffusion in the turbulent medium.
Looking at $\tau_{\rm Fe} / \tau_{\rm crit}$ in Table~\ref{tab:3D_models}, it is clear T42L5.0 will have much less efficient convection in the Fe opacity peak convection zone and thus weaker surface turbulence.
Following the same logic, we expect T19L6.4 to have the strongest convection and fastest velocity fields while T32L5.2 is somewhere in the middle.
Comparing the surface velocities ($v_{\mathrm{r},\,\tau=1}$ and $v_{\bot, \,\tau=1}$) we confirm this to be the case with T19L6.4 having the strongest velocity field and T42L5.0 having the weakest.

\subsection{Photospheric Definition in a Clumpy Surface} \label{subsec:photosphere}

All of our 3D RHD models, independent of $\theta$ and $\phi$, lack the conventional manifestation of a photosphere at a constant radius.
The 3D analogy to a 1D model photosphere would be identifying the location where the angle-averaged optical depth is unity ($r_{\tau = 1}$, see Table~\ref{tab:3D_models}).
However, individual line-of-sight (e.g. along specific $\theta$, $\phi$) optical depths can differ significantly when near-surface convection is vigorous.
Additionally a second photospheric definition, the location where the angle-averaged radiation field follows the Stefan-Boltzmann law ($r_{F_{\rm r} = \sigma T^4}$, see Table~\ref{tab:3D_models}) also disagrees with $r_{\tau = 1}$ by roughly a pressure scale height in the 3D models when turbulence is strong.
Trans-sonic velocity fields, generated by the Fe and He opacity peaks near the surface \citep{Jiang2015, Jiang2018, Schultz2020, Schultz2022}, create large density contrasts ($\rho / \langle \rho \rangle \gtrsim 100$) over a relatively narrow temperature range ($T / \langle T \rangle \lesssim 1$).
As opacity is very sensitive to density in these opacity peak regions, any over-density strongly modify the local heat transport.
The complex topography of the photosphere is also characteristic for constant-tau surfaces for $\tau \gtrsim 100$.

\begin{figure*}
%\vspace*{-2.0 cm}
\centering
 \begin{minipage}[b]{0.45\textwidth}
   \begin{center}
     \includegraphics[width=0.8\textwidth]{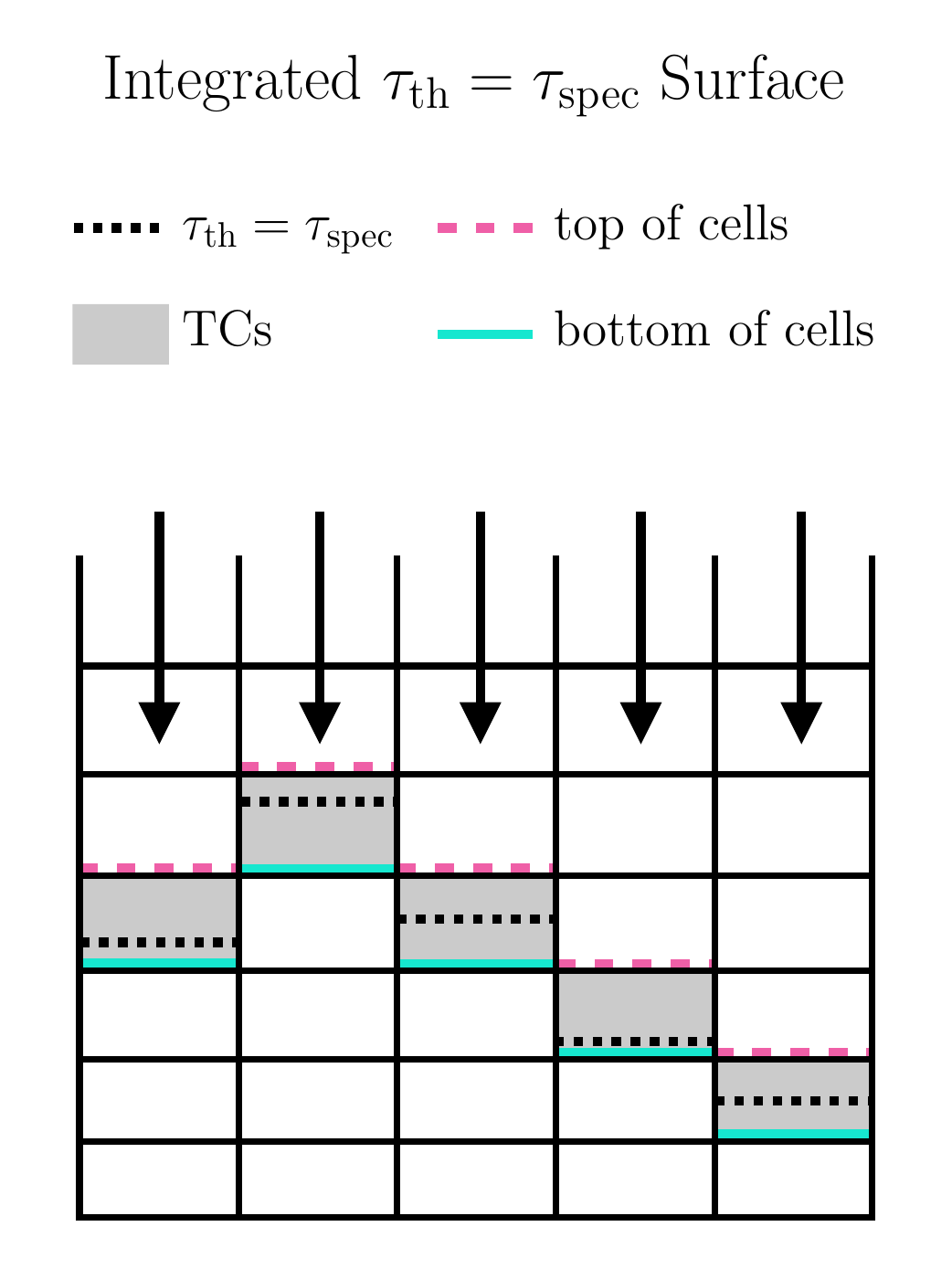} 
     \caption{Schematic of the constant ($\theta,\phi$) integration process for integrated $\tau_{\rm th}=\tau_{\rm spec}$ surfaces, where $\tau_{\rm spec}$ is the desired value for $\tau_{\rm th}$.
     The black dotted line depicts the location along each radial line-of-sight where $\tau_{\rm th}$ is equal to the desired value, while the pink dashed and solid cyan lines show the top and bottom of the cell in which the equality is met respectively.
     The cells themselves, referred to as thermalization cells (TCs), are shaded in grey.}
     \label{fig:isotau_schematic}
    \end{center}
  \end{minipage}
  \hfill
  \begin{minipage}[b]{0.45\textwidth}
    \begin{center}      \includegraphics[width=\textwidth]{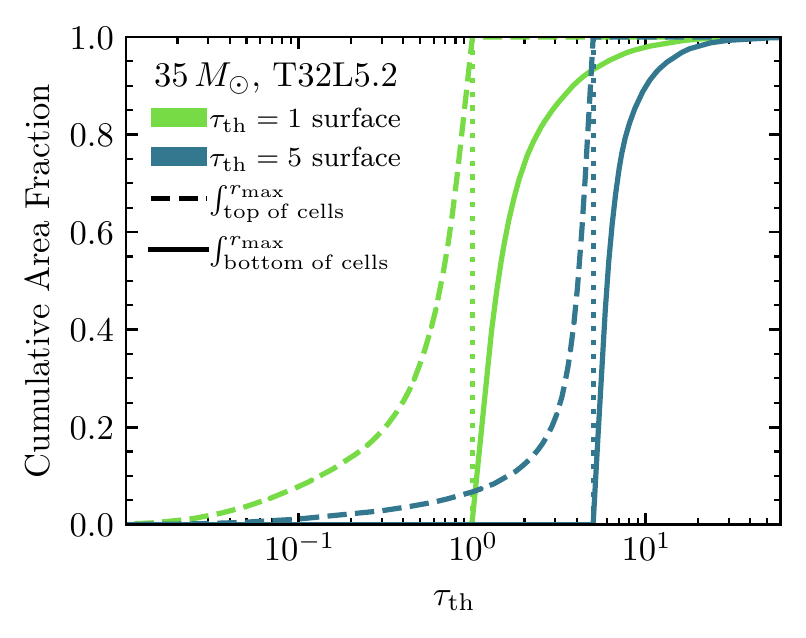} 
      \caption{Variety in definition of integrated $\tau_{\rm th}$ surfaces for T32L5.2 shown as the cumulative area fraction versus thermalization optical depth (see Equation~\ref{eq:tau_th}).
      Linestyles indicate the location of the termination of the optical depth integral, stopping at the top and bottom of the layer cells, $x_{\tau_{\rm th}}$, shown by the dashed and solid lines respectively.
      Vertical dotted lines mark the value of $\tau_{\rm thresh}$ and the colors identify different choices of $\tau_{\rm th}$ for the iso-tau surfaces ($\tau_{\rm th}=1$ and $\tau_{\rm th}=5$ being green and blue respectively).}
      \label{fig:isotau_variety}
    \end{center} 
  \end{minipage}
\end{figure*}

Figure~\ref{fig:rslice} shows $\rho$ variations across radial slices at two choices of integrated volume-weighted, angle-averaged optical depth, $\langle \tau \rangle = 1$ and $\langle \tau \rangle = 10$.
The optical depth is integrated along radial lines-of-sight, lines of constant $(\theta, \phi)$, from the outer boundary of the simulation domain to the cell then angle-averaged to calculate $\langle \tau \rangle$.
In T42L5.0, the density only deviates from the mean by less than 12\% at $\langle \tau \rangle = 1$ as the turbulent velocities are small and subsonic.
In contrast, T32L5.2 and T19L6.4 contain $\rho$ values spanning more than four orders of magnitude at single radial locations resulting from the trans-sonic turbulence at both $\langle \tau \rangle = 1$ and $\langle \tau \rangle = 10$.
As $\langle \tau \rangle$ is approximately the optical depth of the deepest cell, and $\tau_{\rm cell} \approx \kappa \rho \delta r$, the large $\rho$ contrasts with comparatively small changes in $\kappa$ and $\delta r$ result in optical depth variations of $\tau \sim 0.1-1000$ along these radial slices.
Defining a single radius where the angle-averaged optical depth equals a chosen constant is therefore not a reliable approximation for the photosphere. 
Instead, we found that surfaces of constant thermalization optical depth, which can span several scale heights, are the more realistic representation of the local photosphere.

\subsubsection{Thermalization Optical Depth Surfaces}
\label{sec:tau_th}

Identification of the line-emitting region is paramount to spectral synthesis.
Thus instead of looking at $\tau(r) \equiv \int_r^{\infty} \kappa(r')\rho(r') dr'$, where $\kappa(r')$ is the total opacity, we will be using the thermalization optical depth (also known as the effective optical depth),
\begin{equation} \label{eq:tau_th}
\tau_{\rm th} \equiv \int_r^{\infty} \sqrt{\kappa_{\rm a}(r')\left(\kappa_{\rm a}(r')+\kappa_{\rm s}(r') \right)} \rho(r') dr',
\end{equation}
where $\kappa_{\rm a}$ and $\kappa_{\rm s}$ are the absorption and scattering opacities respectively.
This defines the optical depth with regards to creation events (via thermalization) of the photon rather than the location of last scattering.
We can estimate the typical number of emission and absorption events using $N \approx \tau_{\rm th}^2$. 
The condition $\tau_{\rm th}=1$ thus characterizes the regions where the photons are created.

To identify the surface along $\tau_{\rm th}=1$, Equation~\ref{eq:tau_th} is integrated from the outer edge of the simulation domain to the local radius along fixed ($\theta,\phi$) radial lines of sight.
The contribution of the low density floor material ($\rho = 10^{-17}\,$g$\,$cm$^{-3}$) to the optical depth is small ($\tau_{\rm floor} \approx 10^{-5}$) and is ignored.
We identify the outermost cell that contains the location where the integrated $\tau_{\rm th} = 1$, referred to as the thermalization cell (TC).
A schematic diagram of this integration process to accumulate $\tau_{\rm th}$ inwards to a given value, $\tau_{\rm spec}$, as well as the definition of TCs is depicted in Figure~\ref{fig:isotau_schematic}.

\begin{figure*}
%\vspace*{-2.0 cm}
\begin{center}
 \includegraphics[width=0.9\textwidth]{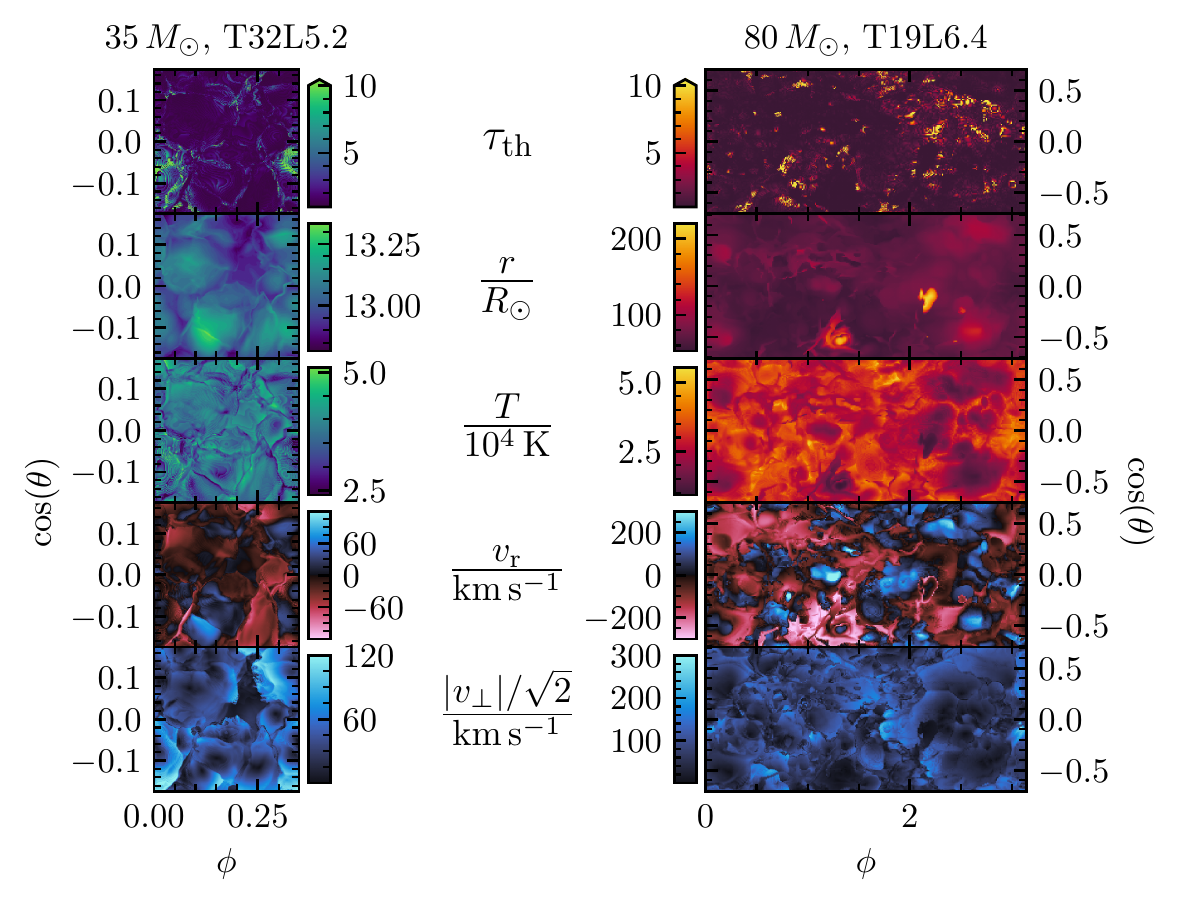} 
%\vspace*{-1.0 cm}
\caption{Properties of models T32L5.2 (left) and T19L6.4 (right) along the integrated $\tau_{\rm th}=1$ surface from the outer boundary of the simulation domain along radial lines of sight.
The thermalization optical depth integrated through the cell ($\tau_{\rm th}$), radius ($r$), temperature, ($T$), radial velocity ($v_{\rm r}$), and tangential velocity magnitude ($|v_{\bot}|/\sqrt{2} = \sqrt{0.5(v_{\theta}^2 + v_{\rm \phi}^2)}$) are plotted from top to bottom.
The colorbars for the tangential velocity panels are identical to the positive (outward velocity) halves of the radial velocity panels.}
 \label{fig:isotau_surfaces}
\end{center}
\end{figure*}

Figure~\ref{fig:isotau_variety} compares the variety of $\tau_{\rm th}$ when integrating to the top (dashed) and bottom (solid) of the TCs in the T32L5.2 model. 
Integrating to the top of the TCs, $10\%$ of lines of sight see significantly lower thermalization optical depths whereas including the entirety of the TCs results in $10\%$ of lines of sight seeing twice the expected $\tau_{\rm th}$.
The variety along the same $\tau_{\rm th}=1$ surfaces in the other two models is comparable to Figure~\ref{fig:isotau_variety}.

The large variety in $\tau_{\rm th}$ of adjacent cells are created by the turbulent structure.
Infalling material, previously ejected from earlier plumes penetrating the surface, collides with rising plumes creating shocks and thus high density contrasts.
These high density contrasts are responsible for the significantly higher $\tau_{\rm th}$ along many of the $\tau_{\rm th}=1$ surfaces (green solid line in Figure~\ref{fig:isotau_variety}).
%In theory, one could use an interpolation method to better estimate the specific radial location and fluid quantities at the iso-tau surface, however as we are interested in post-processing the 3D RHD models to produce spectra, the original grid resolution is sufficient for our purposes.

The spatial variety of integrated $\tau_{\rm th}$ along the bottom of the TCs is shown in the top panel of Figure~\ref{fig:isotau_surfaces}, which also depicts the $r$, $T$, $v_{\rm r}$, and $v_{\bot}$ (in vertical order) for the TCs for T32L5.2 and T19L6.4.
Several distinct plumes can be seen as the radial extent varies in the second-to-top panels with some plumes extending an additional stellar radius in T19L6.4.
The temperature is more or less stratified with higher material being cooler than that beneath it, though the variety in temperature is relatively small.
Looking at $T$ along this surface in T19L6.4 it may be surprising that we find $T_{\rm eff}=19,000\,$K when visually the average temperature along $\tau_{\rm th}=1$ appears closer to $30,000\,$K.
This is due to $\tau_{\rm th}=1$ being below the $\tau = 1$ surface and because the stated $T_{\rm eff}$ is defined by the location where $\langle \tau \rangle = 1$.
The cool, dense plumes shift this photospheric definition further out, reducing the calculated $T_{\rm eff}$.
If instead we use the photospheric definition of $\langle F_{\rm r} \rangle = \sigma \langle T \rangle^4$, we find better agreement (see $T_{\rm eff,\,F_{\rm r}}$ in Table~\ref{tab:3D_models}).
These are just two choices of a photosphere, that are typically consistent in 1D modeling but have drastic differences in 3D models implying more care is needed when defining the emission surface of a 3D clumpy envelope.

In LTE, $T$, $\rho$, and chemical composition determine which lines are excited while velocities determine their resulting profile.
As $T$ varies along the last emitting region, different regions could be exciting different lines.
Furthermore, as temperature is relatively smooth along fixed ($\theta,\phi$), this implies that spectral lines are emitted from a relatively broad region rather than a single radial location.
We also expect significant spectral broadening as both the radial and tangential components of the velocity field are strong ($\gtrsim 100\,$km$\,$s$^{-1}$) with large plumes circulating up and down with substantial tangential motions.
These plumes are a scale height in size setting the scale of the velocity field, which at $\tau=1$, implies a photon mean free path comparable to the turbulent scale.
As the boundary between micro- and macro-turbulent velocities is often identified by the coherence length of the velocity field being larger or smaller than the emitting region, we expect this complex velocity structure to contribute to both the micro- and macro-turbulent broadening.
Both the extended emission region and the turbulent velocity field significantly impact photospheric lines and motivate our spectroscopic analysis.

\section{Post-Processing Using Monte Carlo Methods } \label{sec:rad_trans}

The required computational resources needed for frequency-dependent 3D RHD would be more prohibitively expensive than the grey 3D RHD models.
Instead, a more productive approach to predict observables is to post-process the 3D RHD models with a frequency-dependent radiation transport code. 
To this end, we modify the Monte Carlo radiation transport framework \Sedona, originally developed for synthesizing photometric and spectral observations from stellar transients \citep{Kasen2006}. 

Monte Carlo particles (MCPs), each representing many actual photons, are propagated through a user defined medium, interacting with the matter before reaching the outer boundary of the simulation domain where they are collected and binned to yield a synthetic spectrum.
\Sedona\ computes frequency-dependent opacities for each zone by including the contributions from electron scattering, bound-free/photoionization cross sections \citep{Verner93}, free-free absorption \citep{GM78}, as well as bound-bound/line opacity \citep{Kurucz95,VVF96}.
By default, \Sedona\ incorporates line opacity via the line expansion formalism suitable for the steep velocity gradients of supernovae \citep{EP93}. 
In our work, we remove the line expansion formalism while aggregating the contributions from individual lines to build the opacity table for radiation transport and use atomic data from the \textsc{CMFGEN} database \citep{Hillier2012}.
We adopt solar abundances for both \Athena\ and \Sedona\ models.
We assume LTE in \Sedona\ to compute the ionization fractions and level populations of each atomic species.
%A non-LTE module in \Sedona\ is currently under development and will be used in future work.
As \Sedona\ has been used for numerous frequency dependent  applications \citep[e.g.][]{Kozyreva2020, Tsang2020}, the radiation transport methodology has been extensively tested and compared with other codes, so we focus on describing our substantial modifications that extend \Sedona's application realm to the surfaces of massive stars.

In \Sedona, the MCPs are propagated in Cartesian coordinates, allowing for simple particle propagation during the transport step and the flexibility of utilizing different grid geometries.
As the \Athena\ models used a spherical wedge geometry, which \Sedona\ lacked, the first modification included a 3D spherical wedge geometry with periodic boundary conditions. 
The full sphere geometry was modified to have finite angular extents with periodic boundary conditions in both $\theta$ and $\phi$.
Specifically, when the MCPs encounter ($\theta,\phi$) boundaries, their ($\theta,\phi$) positions are transformed to just inside the opposite boundary, whereas their spherical velocity components ($v_{\rm r}$, $v_{\theta}$, and $v_{\phi}$) are unchanged. 

The second major modification dealt with the large radial extent of the photosphere prevalent in the 3D models.
As shown in Figure~\ref{fig:isotau_surfaces} and discussed in Section ~\ref{sec:grey_models}, the \Athena\ envelope models have very clumpy surfaces with sharp transitions in optical thickness.  
Thus if particles were initialized at a constant radius where all lines of sight had $\tau > 1$, some photons would be propagating through cells with $\tau_{\rm cell} > 100$ causing the MC transport to be computationally prohibitive.
Moreover, radiation transport in such high $\tau_{\rm cell}$ region is tedious as the particles are purely diffusive, and is already well-characterized by the \Athena\ grey model.
Thus, instead of using a constant radius, we defined a custom emitting region (CER) using the TCs described above and avoids the unnecessary computational cost of following long particle histories in diffusive zones.

We initialized the CER using the TCs defined by $\tau_{\rm th}$ surfaces similar to those plotted in Figure~\ref{fig:isotau_surfaces}.
The main difference being that a higher $\tau_{\rm th}$ should be chosen to ensure thermalization before the MCPs reach optically thin material and escape.
Through testing, we found no noticeable difference between $\tau_{\rm th}=5$ and $\tau_{\rm th}=10$ and thus used the $\tau_{\rm th}=5$ surface for the CER in our models.
The co-moving radiative flux, contained in the input model, then defines the initial energy emitted from the desired grid zones of the CER.
The user then defines the total number of particles to emit that is evenly divided among the emitting zones.

\begin{figure*}
%\vspace*{-2.0 cm}
\begin{center}
 \includegraphics[width=0.95\textwidth]{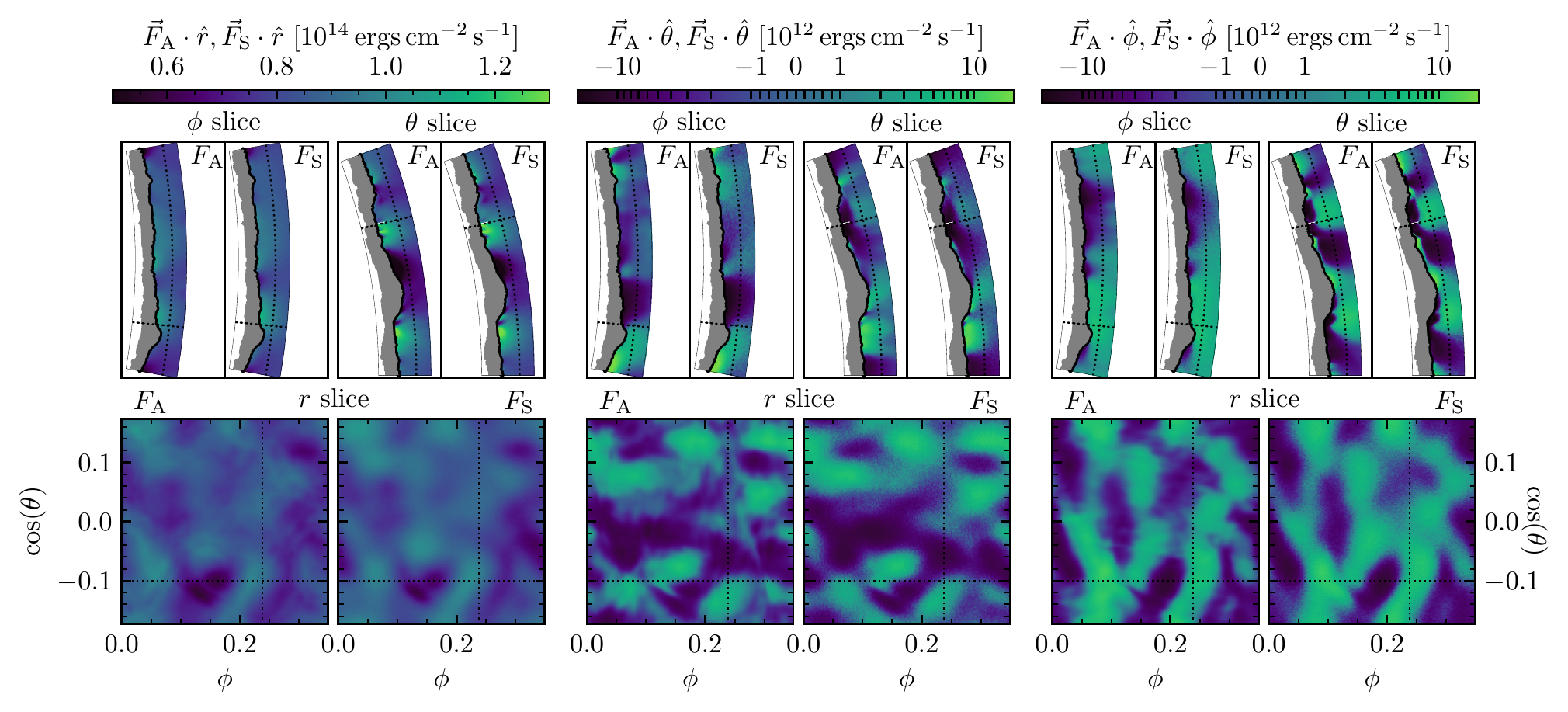} 
%\vspace*{-1.0 cm}
 \cprotect\caption{Side-by-side comparison of radiative flux in grey radiation transport of \Athena, $\vec{F}_{\rm A}$, and \Sedona, $\vec{F}_{\rm S}$ through the same snapshot from T32L5.2.
 The three components of the flux in spherical coordinates are directly compared: radial, longitudinal, and latitudinal from left to right.
 Each component is compared in three slices, one at constant longitude ($\phi$ slice), constant latitude ($\theta$ slice), and constant radius ($r$ slice).
 The grey shaded region shows the extent of the custom emitting region (CER) from the full wedge of the \Sedona\ run with the black line highlighting the outermost extent at $\tau_{\rm th}\approx5$.
 The outer extent of the computational domain is well beyond the photosphere.
 The black dotted lines depict where the slices intersect.
 All panels have unity aspect ratios.}
 \label{fig:grey_F_comp}
\end{center}
\end{figure*}
 
More specifically, the MCPs are randomly generated along the faces through which radiative flux is leaving the cell.
The particles are then isotropically emitted along the direction normal to the cell faces.
To validate this initialization approach, an alternate method was developed in which the MCPs are placed in the center of each zone and isotropically emitted along the direction of the co-moving radiative flux to ensure sufficient angular sampling of the radiation field.
In this work, as particles are only emitted in the optically thick, diffusive regime both initialization methods produce the same results.
As \Athena\ is an Eulerian code with cell centered values, using the first prescription is technically more correct for propagation.
However, as this forces the MCPs to propagate though half the optically thick zone it is noticeably slower than the face emission technique.
Thus for this work we use the face emission method as the results have proven to be identical.

\subsection{Verifying Novel Modifications: Comparing Grey Radiation Transport}
\label{subsec:grey_comp}

The major modifications to \Sedona\ were first tested using a grey radiation transport setup. 
Using $\rho$, $T$, and the grey $\kappa_{\rm Ross}$ from the T32L5.2 \Athena\ model as input, a CER was initialized using the $\tau_{\rm th} = 5$ TCs using the co-moving radiative flux from \Athena.
\Sedona\ then initialized particles and propagated them through the $\tau_{\rm th} < 5$ region of the model, tallying the flux through this region while keeping $T_{\rm gas}$ and $\rho$ fixed.
Figure~\ref{fig:grey_F_comp} compares the spherical components of the co-moving flux above the CER between \Sedona\ and \Athena.
The location of the angular slices were chosen to highlight the agreement across the high dynamic range in radiative flux.
The large scale structures in the radiative flux are well reproduced and the overall radiation transport through the models are consistent.

\Athena\ appears to have some smaller scale structures that is smoothed in the \Sedona\ recreation.
These differences are amplified when comparing the $\theta$ and $\phi$ components of the flux as they are an order of magnitude less than the radial component.
%The discrepancies become more prominent further away from the emission region, developing in the $\approx$10 cells above the outermost emitting cell (shown in the bottom panels of Figure~\ref{fig:grey_F_comp}) and persisting to the outer boundary of the simulation domain.
Regardless of the number of MCPs used in \Sedona, the diffusive nature of radiative flux (highlighted in the $\phi$ slice of the $\theta$ component of the flux in Figure~\ref{fig:grey_F_comp}) remained implying the isotropic emission used in the CER method to likely be the culprit.
Fortunately, the agreement between the grey radiation transport of \Sedona\ and \Athena\ is strong and does not affect the generation of synthetic spectra making it suitable for this work.

It is also interesting to note that the radiation flux at $\tau \geq 1$ is not purely radial.
The large-scale plume structures generate tangential temperature gradients producing diffusive radiative flux in the $\theta$ and $\phi$ directions.
This can be seen in the center and right collections of panels in Figure~\ref{fig:grey_F_comp} where the two components are non-zero even at the emitting layer at $\tau \approx 5$.
These tangential temperature gradients lead to a $\approx 10\%$ deviation from the purely radial gradient used in deriving limb darkening mechanisms at stellar surfaces.
It is unclear how the non-radial temperature gradient will affect limb darkening but we would not expect them to be the same and further investigation is warranted.

\section{Synthesizing Spectra via Post-Processing} \label{sec:post-process}

\begin{figure*}
%\vspace*{-2.0 cm}
\begin{center}
\includegraphics[width=0.6\textwidth]{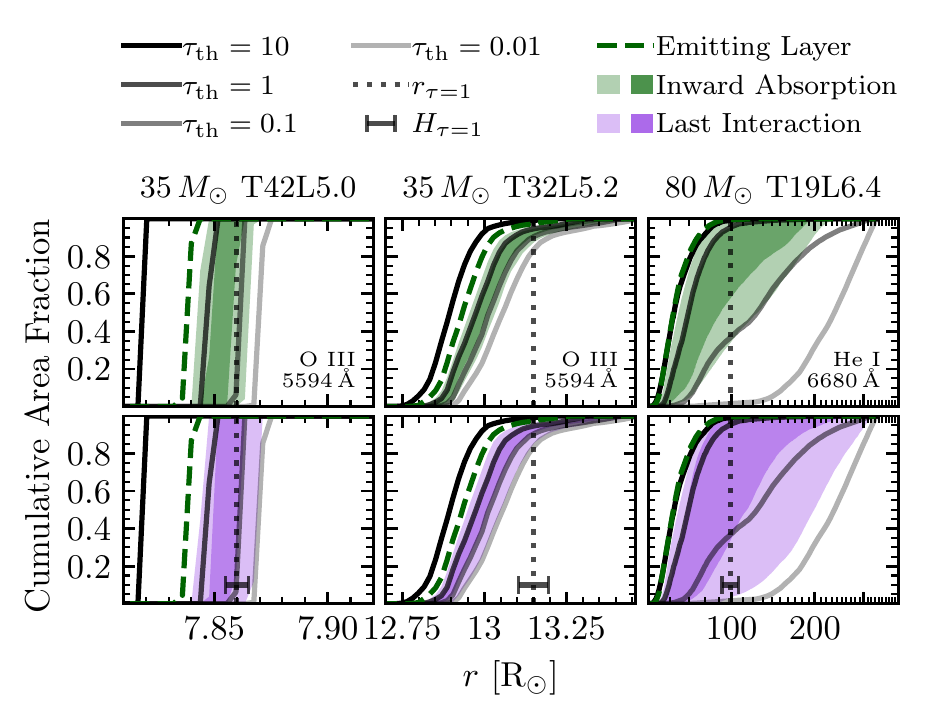} 
%\vspace*{-1.0 cm}
\caption{Comparison of $\tau_{\rm th} = 10,1,0.1,0.01$ surfaces (solid black lines from opaque to transparent respectively) with the regions where Monte Carlo particles (MCPs) are first thermalized when propagating radially inwards (green shaded region in top panels) and the regions where the MCPs last thermalized when propagating outward (purple shaded region in bottom panels) in our 3D models.
Each column is for different 3D models, T42L5.0, T32L5.2, T19L6.4 (left to right respectively) showing individual spectral lines, O III 5594$\,$\AA\ for T42L5.0 and T32L5.2, and He I 6680$\,$\AA\ for T19L6.4.
The two shades of the shaded regions signify the areas where $68\%$ and $95\%$ (lighter and darker respectively) of the MCPs interact. 
The green dashed line shows the furthest inward location where inwardly propagating the MCPs reach, which is also used as the emitting layer for the outward moving the MCPs.
The vertical grey dotted line shows the spherically averaged photosphere, $r_{\tau=1}$ and the grey scale bar shows the spherically averaged scale height at that radius, $H_{\tau=1}$.}
\label{fig:emiss_layers}
\end{center}
\end{figure*}

Performing full frequency-dependent transport through a 3D envelope model, even when post-processing a simulation snapshot, is computationally expensive.
The frequency dependent opacity, $\kappa_{\nu}$, must be calculated for each cell individually based on the local $T$, $\rho$, and chemical composition.
Furthermore, sufficient frequency resolution is needed to resolve the line-width in the co-moving frame, the lower limit of which is set by the thermal velocities of ionic species, $v_{\rm therm} \approx \sqrt{2kT/Am_{\rm p}}$ where $A$ is the atomic number.
For our models, $v_{\rm therm}\sim 1-10\,$km$\,$s$^{-1}$ resulting in a requirement of $\delta\nu/\nu\sim10^{-6}$.
Enforcing the required frequency resolution across a full stellar spectrum becomes extremely memory intensive for the spatial resolution of our models.
Additionally, as $\kappa_{\nu}$ near line centers greatly exceed the opacity of the adjacent continuum, the MCPs tend to be re-emitted near the opacity peaks of spectral lines, increasing the true computational cost over the $\tau_{\rm th}$ estimate.

Therefore, in order to demonstrate the utility of proper spectral modeling, we focus on single, isolated spectral lines.
We chose known photospheric lines (O III at 5594$\,$\AA\ and He I at 6680$\,$\AA) which were sufficiently isolated from neighboring lines to ensure lines would not merge and enough continuum is present for correct normalization.
Additionally, we chose lines whose $\kappa_{\nu}$ peak rose to at least ten times the continuum opacity at each model's photospheric $T$ and $\rho$ to ensure similar, reasonable run-times.
We constrained $\Delta \lambda > 4 (1+v_{\tau=1}/c)*\lambda_0$, where $v_{\tau=1}$ is the maximum photospheric velocity, $c$ is the speed of light, and $\lambda_0$ is the line center.

Because $\kappa_{\nu}$ of the spectral lines, especially near the line centers, dominates the grey continuum opacity, using a $\tau_{\rm th}$ surface calculated using the grey Rosseland mean opacity will result in longer computation times and a far deeper emission location.
However, as we are using a Monte Carlo approach, we can increase the calculation efficiency by identifying the region of last thermalization for the MCPs and using that as a CER.
We determine the layer of last thermalization by initializing MCPs at the outer boundary of the simulation domain and propagate them radially inwards along constant ($\theta,\phi$).
Recording the cells where each MCP first experiences a thermalization event, which is typically a line-absorption event as $\kappa_{\nu} \gg \kappa_{\rm grey}$, provides a good localization to the expected emission region.
Along radial rays with constant ($\theta,\phi)$, cells below the deepest cells in which a thermalization event occurs are used as the CER for spectral synthesis to ensure all MCPs experience at least one thermalization event before escaping.

To further improve computing time we also reduced the angular dimensionality of the 3D RHD models used as input to the spectral synthesis calculation by a factor of 4 (e.g. 256x256 $\rightarrow$ 64x64) by averaging quantities in each cell.
Volume-weighted averages were used for $\rho$ and the radiative energy density, $E_{\rm r}\propto T^4$, while velocities and opacities were mass-weighted and all fluxes were weighted by both opacity and mass.
In preliminary testing this re-binning had no effect on the resulting spectral line profiles when compared to the full resolution.
Combining the lower resolution with the isolation of a single line allowed a full spectral profile of T19L6.4 to be modeled with $200$ frequency bins in $32\,$hours on a single compute node with 40 cores.

The top panels of Figure~\ref{fig:emiss_layers} compare the thermalization locations of inwardly propagating the MCPs to several integrated $\tau_{\rm th}$ surfaces for each model.
Despite being classified as photospheric lines, the majority of observed thermalization locations lie outside the $\tau_{\rm th}=1$ surface and reach out too $\tau_{\rm th}\lesssim0.1$.
Using the inward MCPs to define the CER, the MCPs are propagated outward as detailed in \S~\ref{sec:rad_trans}.
The location of last thermalization for each MCP before it leaves the outer boundary is also recorded and the distribution of these locations are shown by the purple shaded regions in the bottom panels of Figure~\ref{fig:emiss_layers}.
The spatial locations of the green and purple distributions match well, validating the inward propagation method used to approximate the region of last thermalization and ensures each MCP experiences at least 1 thermalization event before escaping.

Figure~\ref{fig:emiss_layers} also highlights the differences in surface variability in the different models.
In T42L5.0, where turbulence is sub-sonic, both the $\tau_{\rm th}$ surfaces and the two shaded regions appear step-like in shape implying the constant radius approximation is acceptable in these cases.
Comparing T42L5.0 to T32L5.2 and T19L6.4, we see much more extended $\tau_{\rm th}$ surfaces and emitting regions, with the more turbulent T19L6.4 spanning nearly an order of magnitude in radius.
Additionally, though the shaded regions do not lie on single $\tau_{\rm th}$ surfaces, their trends do match that of the $\tau_{\rm th}$ surfaces implying the assumption of an emission layer with a finite depth is justified as $\Delta \rho / \rho \ll 1$.

The region of last interaction in all three models can be compared to the local pressure scale height, $H_{\tau=1}$, to estimate the relative importance of micro- and macroturbulence in the spectral broadening.
In T42L5.0, the emitting region spans two local pressure scale heights, which is consistent with the small degree of $\rho$ fluctuations as shown in Figure~\ref{fig:rslice}.
Similarly, the last interaction region in T32L5.2 only spans a single local pressure scale height for most of the surface area, yet covers a much larger radial extent similar to the $\tau_{\rm th}$ surface shown in Figure~\ref{fig:isotau_surfaces}.
The variety of radial location of the emitting region causes the MCPs to thermalize at different $T$, $\rho$, and $v_{\rm r}$, $v_{\bot}$ which should manifest as macroturbulent broadening in the spectral line profile.
The most turbulent model, T19L6.4, displays an emitting region that spans many $H_{\tau=1}$, covering nearly twice the stellar radius in extent.
The turbulent broadening should be dominated by individual plume dynamics as they are expelled from the stellar surface.

\begin{figure*}
%\vspace*{-2.0 cm}
\begin{center}
 \includegraphics[width=0.75\textwidth]{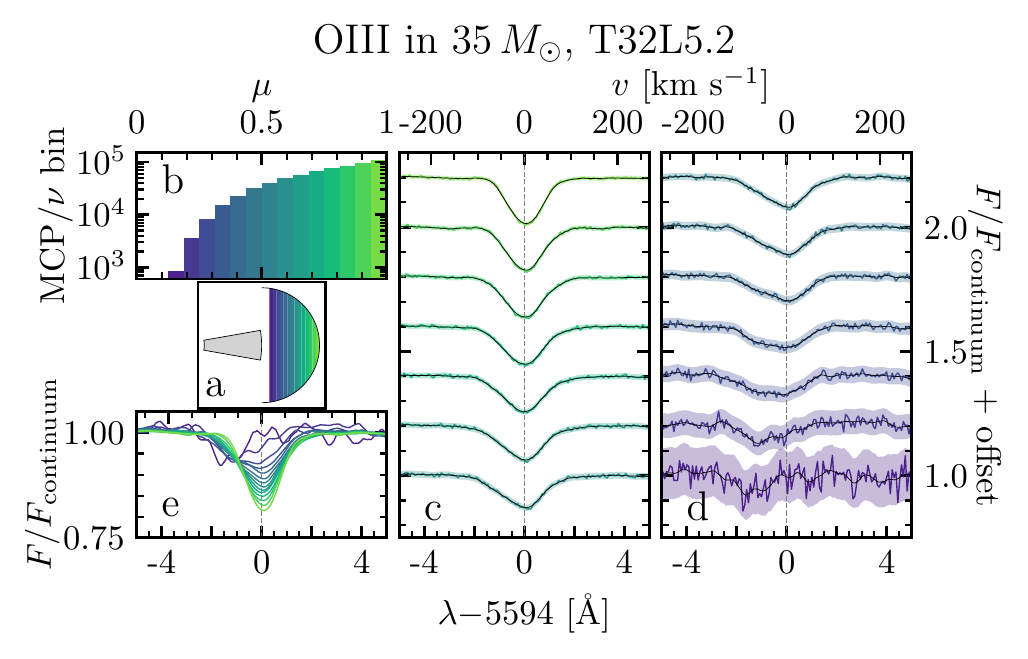} 
%\vspace*{-1.0 cm}
\caption{Analysis of viewing angle dependence in the O III line of T32L5.2.
(a): Schematic showing the simulation domain in grey and the colored areas show the different $\mu \equiv \cos(\theta)$ bins relative the pole of the capturing surface.
In all panels the color represents the given $\mu$ bin shown in this schematic.
(b): Histogram of the median number of MCPs per frequency bin in each $\mu$ bin.
(c) and (d): Normalized spectral lines including a vertical offset.
Colored lines are the raw synthesized spectra which are then smoothed with a Gaussian kernel (black lines).
The colored regions show the $3\sigma$ error from the smoothed spectral line.
(e): Overlay of all smoothed, normalized spectral lines from (c) and (d).}
\label{fig:mu_spec}
\end{center}
\end{figure*}

\section{Variety in Synthetic Spectra of 3D Envelope Models} \label{sec:model_spec}

To model the individual spectral profiles, selected using criteria outlined in \S~\ref{sec:post-process}, we use a frequency range centered on the line center and extending to include three Doppler widths associated with the maximum surface velocity on either side.
Mathematically, the frequencies range from $\nu \in [\nu_0 - 3v_{\rm max}/c, \nu_0 + 3v_{\rm max}/c]$, where $\nu_0$ is the frequency at line center and $v_{\rm max}$ is the maximum magnitude of $v_{{\rm r},\,\tau=1}$ and $v_{\bot,\,\tau=1}$ for each model in Table~\ref{tab:3D_models}.
To ensure adequate sampling of the spectral lines, we divide the frequency range linearly into 200 frequency bins, ensuring that turbulent broadening will be captured by at least ten frequency bins on either side of the line center.

\subsection{Viewing Angle Dependence}
To investigate spectral lines as a function of viewing angle, we divide the outgoing MCPs into 17 $\mu \equiv \cos(\theta)$ bins, where $\theta$ is the angle between the propagation direction of an escaping MCP and the pole.
The pole of each model is defined as the radial unit vector at $\theta = (\theta_{\rm min} + \theta_{\rm max})/2$ and $\phi = (\phi_{\rm min} + \phi_{\rm max})/2$.
Panel (a) of Figure~\ref{fig:mu_spec} shows a schematic of the $\mu$ discretization with color-coded bins, which are are evenly spaced in $\mu$ to standardize the surface area of each bin.
The two most tangential bins are omitted from this analysis (not shown in Figure~\ref{fig:mu_spec}) as few MCPs leave the domain at these small $\mu$. 
As most of the flux is radial, we expect the low-$\mu$ line shapes to be of least interest and thus omit them from the analysis.

The median number of escaping MCPs per frequency bin of outgoing radiation as a function of $\mu$ is shown in panel (b) of Figure~\ref{fig:mu_spec}.
The $\mu=1$ bin receives two orders of magnitude more MCPs than the most tangential bin as a result of the radial flux surpassing the tangential components by an order of magnitude.
This reduction in particle number is also responsible for the increase in spectral noise when going from $\mu = 1$ to $\mu = 0.125$ (green to purple lines in panels (c) and (d) of Figure~\ref{fig:mu_spec}).
This noise is quantified assuming Poisson statistics and using $\sigma = \epsilon \sqrt{N}$ where $\epsilon$ is the average flux carried by each MCP and $N$ is the number of MCPs in each $\nu$ bin.
The shaded regions around each spectral line in panels (c) and (d) of Figure~\ref{fig:mu_spec} show the extent of $3\sigma$ above and below the smoothed spectral line shape (black line).
The smoothed profiles utilize a Gaussian kernel of $0.5-1\,$\AA\ with a $\sigma_{\rm smooth} \approx 0.1\,$\AA.
Panel (e) overlays all the smoothed spectral line profiles for comparison.

\begin{figure*}
%\vspace*{-2.0 cm}
\begin{center}
\includegraphics[width=0.8\textwidth]{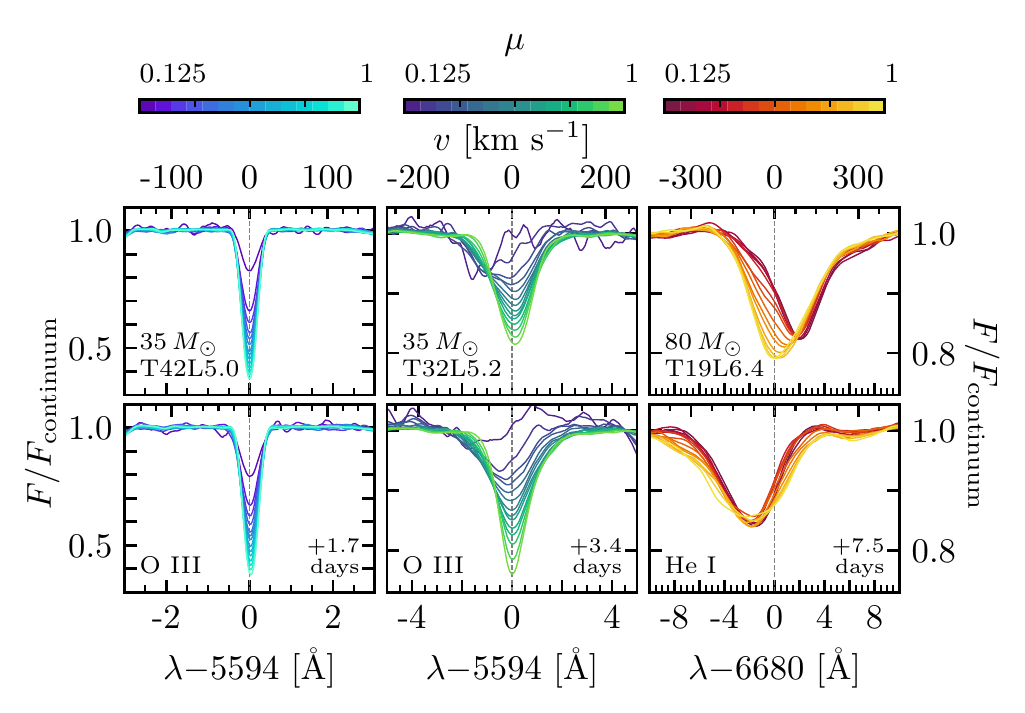} 
%\vspace*{-1.0 cm}
\caption{Comparison of viewing angle (colors) and temporal (top versus bottom panels) dependence on chosen spectral lines in T42L5.0, T32L5.2, and T19L6.4 (left to right).
Each panel is generated equivalently to panel (e) in Figure~\ref{fig:mu_spec}.}
\label{fig:all_spec}
\end{center}
\end{figure*}

The $\mu$ dependence of spectral lines from all three models for two temporal snapshots more than a day apart can be seen in Figure~\ref{fig:all_spec}.
The temporal separation between snapshots is much longer than the eddy turnover times at the Fe opacity peak (1-4 hours) so the surface dynamics of these snapshots should be uncorrelated.
Specifically, these times were chosen to display some of the maximally different spectral profiles that appear in our models.
The three models show very different spectral line shapes as well as temporal line profile variability.

For models T42L5.0 and T32L5.2, the depth of each line appears to change as a function of $\mu$.
Line profiles at lower $\mu$, generated from MCPs with more tangential trajectories, appear to have shallower depths, and thus smaller equivalent widths.
These decreases in the $\mu$ dependence of the absorption depth are likely due to a combination of limb darkening and the finite wedge geometry of the simulation domain.
%As the emitting region in T42L5.0 is similar to a spherical shell, only the geometry of the simulation domain could explain this $\mu$ dependence.
This is further confirmed by the similar $\mu$ dependence of T32L5.2 (also with a small simulation size) and the lack of this $\mu$ dependence in T19L6.4, which covers much more of the stellar surface.
However in T32L5.2, the low $\mu$ trajectories are also broadened and less symmetric as the emitting region is more complex with dynamic topography and velocity fields arising from the stronger turbulent plume motions.
The additional broadening is a result of $v_{{\bot},\,\tau=1} > v_{{\rm r},\,\tau=1}$ (see Table~\ref{tab:3D_models}), which broadens the more tangential trajectories of low $\mu$ bins.
The asymmetry originates from the development of plume structures.
As individual plumes are now separated from each other, their bulk motion affects the spectral line shape by Doppler shifting part of the profiles.
As the individual plumes do not cover a large area fraction, most of the effects cancel resulting in the high $\mu$ peaks being nearly centered on the line center.
However, individual plumes with large tangential velocities cause the slight shifts away from the line center in the low $\mu$ profiles.

The $\mu$ dependence is starkly different in T19L6.4 than the previous two models.
The stellar model is cooler resulting in a larger scale height, and therefore larger plumes.
The motion of individual plumes, which cover a significant fraction of the stellar surface, significantly alter the spectral profiles.
For example, the redshift seen in the dark red lines, low $\mu$ profiles, in the upper right panel of Figure~\ref{fig:all_spec} are the result of plumes previously launched from the stellar surface falling back down.
In contrast, the central regions (yellow lines in the same panel) do not exhibit a dominant plume motion and thus show no bulk Doppler shift.
In the second snapshot (bottom right panel), all the $\mu$ profiles are blue shifted as several uncorrelated plumes are being launched in different directions.
The temporal changes in spectral shape should be observable in some of the models and are explored further in the next section when more realistic synthetic profiles are generated.

\subsection{Synthetic Spectral Profiles}
Our 3D RHD models only cover a patch of the stellar surface and especially in T42L5.0 and T32L5.2 the small patch size may affect the resulting spectral shapes.
One way to produce more realistic spectral profiles would be to tessellate uncorrelated simulation domains together to generate a hemispherical model.
Post-processing this hemispherical model would produce more realistic spectral profiles.
The computational memory overhead of such an analysis depends on the atmospheric $H/r$ value, and for our current models would be computationally challenging, but may be explored in future work.
Another solution is to use the viewing angle dependence to perform a weighted average of uncorrelated snapshots.
In this subsection we present approximate spectral profiles based on combining the $\mu$ binned spectral lines from the temporally uncorrelated snapshots shown in Figure~\ref{fig:all_spec}.
Because we are only using two snapshots, we refrain from performing detailed quantitative spectral line analysis for the synthesized profiles.
However, an approximate quantitative analysis of the different broadening mechanisms is provided below.

\begin{figure*}
%\vspace*{-2.0 cm}
\begin{center}
\includegraphics[width=0.8\textwidth]{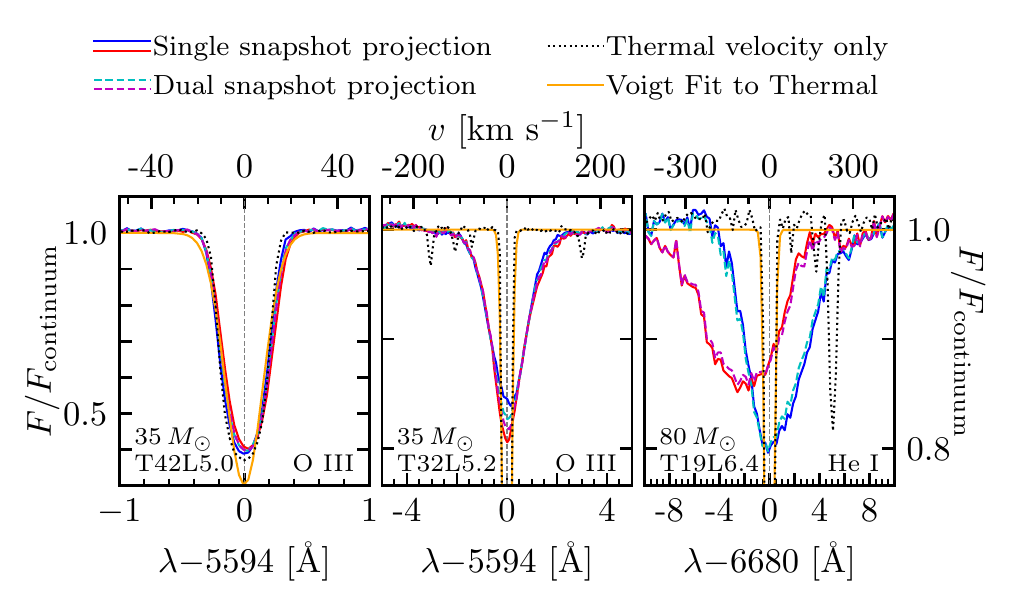} 
%\vspace*{-1.0 cm}
\caption{
Synthetic spectral lines generated from combining the $\mu$ dependent components shown in Figure~\ref{fig:all_spec}.
Three combinations were chosen: the solid red and blue lines show single snapshots estimates, the dashed cyan and magenta lines show when alternating $\mu$ bins were used from both of the temporal snapshots, and the black dotted line shows a single snapshot with only the thermal velocity used.
The thermal velocity profile was fit to a Voigt profile (solid orange line).
}
\label{fig:proj_spec}
\end{center}
\end{figure*}

As the narrow models (T42L5.0 and T32L5.2) are small (covering only a 50th of the stellar surface), we perform a weighted sum of the spectral flux from individual snapshots using each $\mu$ bin's projected area on the stellar disk as weights.
Dividing by the mean of the resulting continuum produces the solid red and blue lines in the left and center panels of Figure~\ref{fig:proj_spec}.
The same procedure is performed for T19L6.4 however as the original angular coverage is much larger (covering $\approx 80\%$ of the stellar disk), the $\mu=1$ bin is taken for the whole area of the simulation domain and weighted by its projected area on the stellar disk.
The remaining area of the stellar disk is added to the weighted sum using the same procedure as above and the normalized spectral profiles are shown by solid red and blue lines in right panel of Figure~\ref{fig:proj_spec}.
In an attempt to represent what an uncorrelated surface may look like, we perform a weighted average using alternating $\mu$ bins from the two available snapshots. 
Shown by the dashed lines in Figure~\ref{fig:proj_spec}, the similar colors represent which snapshot's $\mu=1$ bin was used (e.g. the cyan dashed line used the same central bin as the blue solid line).
Ideally, we would have used many temporally uncorrelated snapshots in the projection, however as computational resources are limited, we leave this as a future exercise.

Taking these simple synthetic spectral profiles as truth, we predict all three models to show some temporal variability.
The variability in T42L5.0 would be very challenging to observe, requiring a $R \sim 200,000$ spectrometer to measure the slight line center shift and very high signal to noise to catch the change in the depth of the absorption line.
Model T32L5.2 would show signs of variability with only a $R\sim 20,000$ spectrograph with sufficient signal to noise to resolve the change in line shape.
The temporal variability of the third model, T19L6.4, should be easy to resolve with many spectrometers.
All small fluctuations present in the colored spectral profiles are due to the Poisson noise from the finite number of MCPs and are compounded by the fact only two uncorrelated snapshots were used.
If more snapshots and MCPs were used, we would expect significantly smoother lines.
Additionally, the inclusion of more uncorrelated snapshots in the projection process may reduce variability, further constraining the challenge of observing the temporal line variability.

\subsection{Thermal Profile Comparison}
To investigate how the bulk velocity field of the cells affects the spectral lines, additional calculations were run with all the cell velocities set to zero to generate spectral profiles resulting only from thermal broadening.
Shown by the black dotted lines in Figure~\ref{fig:proj_spec}, all the models display noticeable differences from the inclusion of velocity fields.
As expected, the cell velocities broaden the profiles and decrease their absorption amplitude, providing strong evidence for the presence of macroturbulence.

To quantify their widths, each thermal profile was fit to a Voigt profile (orange solid lines in Figure~\ref{fig:proj_spec}).
As can be seen in the left panel, but is also present in the other two panels, the Voigt profiles does not fit the full thermal profile very well: the peak is too narrow and the tails are too wide.
The discrepancies arise despite the fact that the fit is dominated by the Gaussian component of the Voigt profile, making the peak as wide and the tails as narrow as possible.
The lack of a Lorentzian component is not surprising as \Sedona\ does not include natural broadening in the calculation of $\kappa_{\rm nu}$.
However at the high temperatures of these massive star surfaces and for our chosen spectral lines, the intrinsic widths should be small compared to the thermal broadening, justifying this assumption.

Despite the shape mismatch between the thermal profiles, the widths of the fits appear to be approximately correct, especially at the full width half minimum (FWHM), implying the standard deviation of the distribution could be used to estimate a thermal velocity using $v_{\rm fit} = \sigma c / \nu_{0}$, where $\nu_0$ is the frequency at line center.
However, $v_{\rm fit} \approx 1.3\,v_{\rm therm}$ when the thermal velocity is calculated with the temperatures of the emitting region, using $T = T_{\rm eff}$ (e.g. for T32L5.2: $v_{\rm fit} = 7.5\,$km$\,$s$^{-1}$ and $v_{\rm therm} = 5.7\,$km$\,$s$^{-1}$).
This difference cannot be explained by microturbulence as $v_{\rm therm}$ is the only velocity in the calculation of the thermal line profiles.
Thus the discrepancies are either due to the range of $T$ across the emitting region (which vary by $\gtrsim 50\%$ about the mean) or imperfections in the projection method.

The four smaller troughs of the black dashed line in the center panel and the larger valley to the right of line center in the right panel of Figure~\ref{fig:proj_spec} are other spectral lines with less amplitude than the main profile being investigated.
In the central panel, they do not impact the line shape significantly, however in the right panel of Figure~\ref{fig:all_spec} it is clear the additional absorption line affects the line shape in the right tail.
However, as the auxiliary line is nearly $5\,$\AA\ ($250\,$km$\,$s$^{-1}$) away from the main line center and is four times weaker than the He I line we are investigating, it does not affect our analysis.

\subsection{The Effect of Micro- and Macroturbulence}
All three panels of Figure~\ref{fig:proj_spec} clearly show additional broadening beyond thermal broadening.
As there is no rotation in the models, this additional broadening can only be attributed to the impact of the envelopes' velocity field, resulting in $\xi$ and $v_{\rm macro}$.
However it is difficult to determine the individual impact of $\xi$ or $v_{\rm macro}$ on the broadening without a full quantitative analysis.
The simplest distinction could be made by comparing the equivalent widths of the thermal profiles to the full velocity field lines.
If the equivalent width was conserved, $v_{\rm macro}$ is likely the dominant factor and $\xi < v_{\rm therm}$ \citep{Gray2005}.
Conversely, if the equivalent width changes, $\xi > v_{\rm therm}$ and microturbulence is likely playing a dominant role, making it hard to quantify $v_{\rm macro}$.

Unfortunately due to the noise present in the spectral line shapes, it is hard to compare the equivalent widths between the lines without pushing to more MCPs or using more temporally uncorrelated patches.
However, comparing the FWHM of each line can tell us about the dominant velocity.
For T42L5.0, the FWHM is $ \approx 20\,$km$\,$s$^{-1}$ which is consistent with thermal broadening as the FWHM for a Gaussian $ \approx 2.4\sigma$.
Visually however, it is clear the tails of the synthetic profiles in the left panel of Figure~\ref{fig:proj_spec} are wider than the thermal profile suggesting the additional presence of macroturbulent broadening.
For T32L5.2 the FWHM velocity of $\approx 80\,$km$\,$s$^{-1}$ is comparable to the surface velocities, $v_{{\rm r},\,\tau=1}$ and $v_{{\bot},\,\tau=1}$ in Table~\ref{tab:3D_models}, suggesting $\xi$ and $v_{\rm macro}$ are important.
Due to the trumpet like shape of the profiles, we suspect $v_{\rm macro}$ to be more dominant though thorough fitting is required to ascertain.
The FWHM velocity of T19L6.4 is $\approx 275\,$km$\,$s$^{-1}$ and is nearly double the surface velocities in Table~\ref{tab:3D_models}.
This suggests both $\xi$ and $v_{\rm macro}$ are important in this very turbulent model and more investigation of the broadening mechanisms is warranted.

Qualitatively, our calculated spectral profile for the OIII line in T32L5.2 strongly resembles observed spectral profiles seen in \cite{SimonDiaz2014}.
As our models contain no rotation, the $v\sin i$ values do not agree between our calculations and observations.
However the observed macroturbulence velocities agree quite well with the calculated FWHM and thus the turbulent velocities in the simulation.
Unfortunately, no observations of stars similar to T42L5.0 or T19L6.4 have been carried out.
Future spectroscopic surveys \citep[e.g.][]{Bowman2022} will observe stars that could be directly compared to our other models. 

\section{Conclusion and Future Work} \label{sec:conclusion}

Three dimensional spectroscopic analysis of hot massive star surfaces is now possible using the \Sedona\ Monte Carlo radiation transport code to post-process 3D RHD, LTE \Athena\ models.
The turbulence excited by near-surface convection zones, predominantly those generated from the Fe and He opacity peak, create large $\rho$ and velocity contrasts that persist through the $\tau=1$ surface.
Because of the inherent plume structure of the turbulent motions and many order of magnitude density contrasts, the $\tau=1$ surface is not uniform in radius but rather spans several pressure scale heights resulting in factor of two variations in $T$.
The $T$, $\rho$, and $v$ fluctuations impact photospheric line broadening implying observed spectroscopic broadening could be used to determine the presence and strength of photospheric turbulence.

The viewing angle dependence appears to deviate from classical limb darkening predictions partially due to the small simulation domain but also due to the tangential temperature gradients generated from the dynamic plume structures at the surface.
Additionally synthesized spectral lines of our models predict observable spectral line variability in Hertzsprung gap stars on several day timescales.
Unfortunately, these calculations only use a small simulation domain and future calculations which mimic the full observed hemisphere of the star are needed for stronger predictions. 

Though our calculated line profiles clearly exhibit the large velocities realized in our simulations, it remains to be seen whether they are well represented by the commonly used macroturbulent fitting approach of \cite{Gray2005}. 
For many OB stars, the only inference regarding rotation comes from broad spectral lines that yield a $v\sin i$ based on the assumption that rotation is a dominant broadening mechanism carried over from analyses of lower mass stars.
Our work here may confound such inferences, as the broadening we see from turbulence alone may well exceed that from any rotation in many instances, especially stars viewed along its rotation axis.
Indeed, the lack of reports of low $v\sin i$ with large $v_{\rm macro}$ speaks to this possible challenge to the rotational interpretations.

Once spectral lines are synthesized from hemispherical stellar surfaces, we could use the same methodology as observers to fit rotational and turbulent velocities to the spectral lines.
These spectroscopic measurements could then be compared with fluid velocities in the 3D models.
Many spectral lines from single ions could be synthesized and combined to make a curve of growth allowing $\xi$ to be identified and related to intrinsic turbulent velocities of the plasma.
With microturbulence quantified, $v_{\rm macro}$ could be identified by calculating the additional broadening needed to fit the spectral line profiles.
Lastly though the models lack rotation, we could set a bulk rotational velocity distribution to a hemispherical model and begin to probe the interplay between $v\sin i$, $\xi$ and $v_{\rm macro}$ in massive star surfaces.
Better understanding of the interplay between velocity fields in massive star surfaces could lead to a better theoretical understanding and a more physically motivated explanation of spectral line fitting parameters.

The methodology presented in this work is not specific to our 3D RHD models of hot massive stars.
Recently novel 3D red supergiant envelope models presented in \cite{Goldberg2022} could be analyzed to better understand how larger plume structures affect recent spectroscopic measurements \citep[e.g.][]{Guerco2022}.
Though these models were also run with \Athena\, any 3D model could be used in this method.
Additionally these methods are not specific to optical lines.
Any line that is emitted near the stellar surface where LTE assumptions are still decent could be analyzed with these methods, though non-LTE radiation transport is currently being investigated.
With these new methodologies, a comprehensive spectroscopic analysis of massive star surfaces is within reach.

%%%%%%%%%%%%%%%%%%%%%%%%%%%%%%%%%%%%%%%%%%%%%%%%%%%%%%%%%%%%%%%%%%%%%%%%%%%%%%%%%%%%%%%%%%%%%%%%%%%%%%%%%%%%%%%%%%%%%%%%%
\section*{Acknowledgements}
We thank Jared Goldberg, Jorick Vink, Sergio Sim\'{o}n-D\'{i}az, and Sunny Wong for many helpful conversations and comments.
This research was supported in part by the NASA ATP grant ATP-80NSSC18K0560, by the National Science Foundation through grant PHY 17-48958 at the KITP.
Resources supporting this work were also provided by the NASA High-End Computing (HEC) programme through the NASA Advanced Supercomputing (NAS) Division at Ames Research Center. 
The founding of this research began at a meeting funded by the Gordon and Betty Moore Foundation through Grant GBMF5076.
We acknowledge support from the Center for Scientific Computing from the CNSI, MRL: an NSF MRSEC (DMR-1720256) and NSF CNS-1725797.
The Flatiron Institute is supported by the Simons Foundation.

%\bibliography{sample631}{}
\bibliography{test}{}
\bibliographystyle{aasjournal}

\end{CJK*}
\end{document}